\documentclass[12pt,reqno]{amsart}
\usepackage[dvips]{graphicx}
\usepackage{amsmath,amssymb,latexsym,indentfirst,times}

\numberwithin{equation}{section}
\newtheorem{theorem}{Theorem}[section]

\newtheorem{proposition}[theorem]{Proposition}
\newtheorem{lemma}[theorem]{Lemma}
\newtheorem{corollary}[theorem]{Corollary}

\newtheorem{remark}[theorem]{Remark}

\newcommand{\alphaparenlist}{% changes enumerate 1st level to (a)...(z)
  \renewcommand{\theenumi}{\alph{enumi}}
  \renewcommand{\labelenumi}{{\rm (\theenumi)}}
}
\newcommand{\alphaprimeparenlist}{% changes enumerate 1st level to (a)...(z)
  \renewcommand{\theenumi}{\alph{enumi}}
  \renewcommand{\labelenumi}{{\rm (\theenumi')}}
}

\newcommand{\mf}{\mathfrak}
\newcommand{\mb}{\mathbb}

\newcommand{\ZZ}{\mathbb{Z}}
\title{Principally specialized characters of $\widehat{\mf{sl}}(m|1)$-modules
}
\author{{\sc Takuya Murakami}}
\date{\today}
\address{Graduate School of Mathematics, Kyushu University 33, Fukuoka,
812-8581, Japan}
\email{ma299040@math.kyushu-u.ac.jp}
\begin{document}

\maketitle
\setcounter{section}{-1}
%%%%%%%%%%%%%%%%%%%%%%%%%%%%%%%%%%%%%%%%%%%%%%%%%%%%%%%%%%%%%%%%%%%%%%%%%
%%%%%%%%%%%%%%%%%%%%%%%%%%%%%%%%%%%%%%%%%%%%%%%%%%%%%%%%%%%%%%%%%%%%%%%%%
%%%%%%%%%%%%%%%%%%%%%%%%%%%%%%%%%%%%%%%%%%%%%%%%%%%%%%%%%%%%%%%%%%%%%%%%%
\begin{abstract}
In this paper, we calculate a series of principally specialized
 characters of the $\widehat{\mf{sl}}(m|1)$-modules of level 1. In
 particular, we show that the principally specialized characters of the
 basic modules $L(\Lambda_0)$ is expressed as an
 infinite product. In addition, we deduce the specialized character
 formula of ``quasiparticle'' type.
\end{abstract}

%%%%%%%%%%%%%%%%%%%%%%%%%%%%%%%%%%%%%%%%%%%%%%%%%%%%%%%%%%%%%%%%%%%%%%%%%
%%%%%%%%%%%%%%%%%%%%%%%%%%%%%%%%%%%%%%%%%%%%%%%%%%%%%%%%%%%%%%%%%%%%%%%%%
%%%%%%%%%%%%%%%%%%%%%%%%%%%%%%%%%%%%%%%%%%%%%%%%%%%%%%%%%%%%%%%%%%%%%%%%%
\section{Introduction}
\label{sec:0}

Character formulas of basic representation of $\widehat{\mf{sl}}(m|n)$ are
given in \cite{KW2}, based on their explicit construction in terms of
bosonic and fermionic fields.

In this paper, we calculate the principally specialized characters 
of some $\widehat{\mf{sl}}(m|1)$-modules.
In \S\ref{sec:1}, we describe that the principally specialized characters of the
 basic $\widehat{\mf{sl}}(m|1)$-modules $L(\Lambda_0)$ is expressed as an
 infinite product.
In \S\ref{sec:2}, we deduce the specialized character formula of
``quasiparticle'' type.

We follow notation and terminologies from \cite{KW2} without repeating
their explanation.

%%%%%%%%%%%%%%%%%%%%%%%%%%%%%%%%%%%%%%%%%%%%%%%%%%%%%%%%%%%%%%%%%%%%%%%%%
%%%%%%%%%%%%%%%%%%%%%%%%%%%%%%%%%%%%%%%%%%%%%%%%%%%%%%%%%%%%%%%%%%%%%%%%%
%%%%%%%%%%%%%%%%%%%%%%%%%%%%%%%%%%%%%%%%%%%%%%%%%%%%%%%%%%%%%%%%%%%%%%%%%
\section{Specialized character formula for some series of $\widehat{\mf{sl}}(m|1)$-modules}
\label{sec:1}

Throughout this paper, we assume that $m\geq2$ and 
let $\alpha_0,\ldots ,\alpha_m$ denote the set of simple roots for
$\widehat{\mf{sl}}(m|1)$ where $\alpha_0$ and $\alpha_m$ are odd and
$\alpha_i (i=1,\ldots ,m-1)$ are even.
Provided that all $s_i$ are positive integers, the sequence $s=(s_0,\ldots ,s_m)$ defines a
homomorphism $\mathcal{F}_s:\mb
C[[e^{-\alpha_0},\ldots,e^{-\alpha_m}]]\longrightarrow\mb C[[q]]$ by
$\mathcal{F}_s(e^{-\alpha_i})=q^{s_i}(i=0,\ldots,m)$, 
called the \emph{specialization of type} $s$.
In this paper, we consider the specialization of type
$s=(1,\ldots,1,0)$ which makes sense for the characters of integrable
representations, and we write simply $\mathcal{F}$ for this
specialization $\mathcal{F}_s$ when no confusion can arise.

Since 
the set of simple roots for the even part of $\widehat{\mf{gl}}(m|1)$ is 
\begin{displaymath}
\widehat{\Pi}' =  \{ \alpha'_0 = \alpha_0+\alpha_m ,
\alpha_1 , \ldots , \alpha_{m-1}\},
\end{displaymath}
this specialization is the principal specialization with respect to the
even part of $\widehat{\mf{gl}}(m|1)$; namely
\[
\mathcal{F}(e^{-\alpha'_0})=\mathcal{F}(e^{-\alpha_1})=\cdots=\mathcal{F}(e^{-\alpha_{m-1}})=q.
\]

We recall the Fock space and its charge decomposition:
\[
 F=\oplus_{s\in \mb Z}F_s
\]
from \S3 of \cite{KW2}.
\begin{lemma}
\label{lem:1.1}
Let $m\geq 2, s\in \mb Z$. Then we have{\rm :}
\alphaparenlist
\begin{enumerate}
\item $\displaystyle q^{\frac{sm}2}\mathcal{F}(e^{-\Lambda_0}ch F_s)+
q^{-\frac{sm}2}\mathcal{F}(e^{-\Lambda_0}ch F_{-s})=\frac{2\prod^{\infty}_{i=1}(1+q^i)^2}{\varphi(q^m)^2}$.
\item $\mathcal{F}(e^{-\Lambda_0}ch F_s)=\mathcal{F}(e^{-\Lambda_0}ch F_{m-1-s})$.
\end{enumerate}
Here and further $\varphi(q)=\prod^{\infty}_{j=1}(1-q^j)$.
\end{lemma}
\begin{proof} \quad 
In the case $\widehat{\mf{gl}}(m|1)$, the
formula (3.15) in \cite{KW2} gives the following:
\begin{equation}
ch F=e^{\Lambda_0}\prod^{\infty}_{k=1}
\frac{\prod^m_{i=1}(1+ze^{\epsilon_i-(k-\frac12)\delta})
(1+z^{-1}e^{-\epsilon_i-(k-\frac12)\delta})}
{(1-ze^{\epsilon_{m+1}-(k-\frac12)\delta})
(1-z^{-1}e^{-\epsilon_{m+1}-(k-\frac12)\delta})}\ ,
\end{equation}
where $z$ is the ``charge'' variable.

Since $\alpha_0=\delta-\epsilon_1+\epsilon_{m+1}$ and
$\alpha_i=\epsilon_i-\epsilon_{i+1}, \ (i=1,\ldots ,m)$, our principal
specialization $\mathcal{F}=\mathcal{F}_s$ is written in terms of
$e^{-\epsilon_i}$ as follows:
\begin{equation}
\label{eq:1.2}
\mathcal{F}(e^{-\epsilon_i})=q^{m-i}(i=1,\ldots,m) ,
\quad \mathcal{F}(e^{-\epsilon_{m+1}})=1.
\end{equation}

Thus, we obtain
\begin{align}
\mathcal{F}(e^{-\Lambda_0}ch F)
&=\prod^{\infty}_{k=1}\frac{\prod^m_{i=1}
(1+zq^{i-m}q^{m(k-\frac12)})(1+z^{-1}q^{m-i}q^{m(k-\frac12)})}
{(1-zq^{m(k-\frac12)})(1-z^{-1}q^{m(k-\frac12)})}
\nonumber \\
&=\prod^{\infty}_{k=1}
\frac{(1+zq^{k-\frac m2})(1+z^{-1}q^{k-1+\frac m2})}
{(1-zq^{m(k-\frac12)})(1-z^{-1}q^{m(k-\frac12)})}
\nonumber \\
&=\prod^{\infty}_{k=1}
\frac{(1+(zq^{\frac{1-m}2})q^{k-\frac12})
(1+(z^{-1}q^{\frac{m-1}2})q^{k-\frac12})}
{(1+(-z^{-1})q^{m(k-\frac12)})(1+(-z)q^{m(k-\frac12)})}\ .
\label{eq:1.3}
\end{align}

In order to compute the coefficient of $z^s$, 
we use the Jacobi triple product identity:
\begin{equation}
\label{eq:1.4}
\prod^{\infty}_{n=1}(1+zq^{n-\frac12})(1+z^{-1}q^{n-\frac12})
=\frac1{\varphi(q)}\sum_{j \in {\mb Z}}z^jq^{\frac12j^2},
\end{equation}
and also the following well-known identity (see (5.26) in \cite{KP} and \S5.8
in \cite{K4}):
\begin{eqnarray}
  \lefteqn{\hspace{-6ex}\prod^{\infty}_{k=1} (1+zq^{k-\frac{1}{2}})^{-1}
  (1+z^{-1}q^{k-\frac{1}{2}})^{-1}=\frac{1}{\varphi (q)^2}
  \sum_{m\in\ZZ}(-1)^m
  \frac{q^{\frac{1}{2}m(m+1)}}{1+zq^{m+\frac{1}{2}}}}
\nonumber \\
  &=& \varphi (q)^{-2} \left( \sum_{m,k \geq 0} -\sum_{m,k<0} \right) 
  ((-1)^{m+k} z^k q^{\frac{1}{2}m(m+1)+(m+\frac{1}{2})k}) \, .
\label{eq:1.5}
\end{eqnarray}
Replacing $z$ by $zq^{\frac{1-m}2}$ in (\ref{eq:1.4}), $z$ by 
$-z^{-1}$ and $q$ by $q^m$ in (\ref{eq:1.5}),
we rewrite the right side of (\ref{eq:1.3}) by
\begin{equation*}
\mathcal{F}(e^{-\Lambda_0}ch F)=
\frac1{\varphi(q)\varphi(q^m)^2}
\sum_{k\in{\mb Z}}(\sum_{a,p\geq0}-\sum_{a,p<0})
(-1)^az^{k-p}q^{\frac12k(k+1)-\frac{km}2+\frac m2a(a+1)+m(a+\frac12)p}
\end{equation*}
and thus we have
\begin{align}
\label{eq:1.6}
\mathcal{F}(e^{-\Lambda_0}ch F_s)
&=\frac1{\varphi(q)\varphi(q^m)^2}
(\sum_{a,p\geq0}-\sum_{a,p<0})(-1)^aq^{\frac12(p+s)(p+s+1)
-\frac{sm}2+\frac m2a(a+1)+map}.
\end{align}

For convenience, we introduce the following functions:
\begin{align}
f_s(a,p)&=(-1)^aq^{\frac12(p+s)(p+s+1)-\frac{sm}2+\frac m2a(a+1)+map},
\nonumber \\
h_s&=(\sum_{a,p\geq0}-\sum_{a,p<0})f_s(a,p).
\label{eq:1.7}
\end{align}

To prove this lemma, it is sufficient to prove the following two equations:
\alphaprimeparenlist
\begin{enumerate}
\item
     $q^{\frac{sm}2}h_s+q^{-\frac{sm}2}h_{-s}=2\varphi(q)\prod^{\infty}_{i=1}(1+q^i)^2$.
\item $h_s=h_{m-1-s}$.
\end{enumerate}

First we shall prove (a').\\
Since $f_s(-a,-p-1)=q^{-sm}f_{-s}(a,p)$, we have 
\begin{align*}
\sum_{a>0 \atop p\geq0}f_s(a,p)&=\sum_{a,p<0}f_s(-a,-p-1)
=q^{-sm}\sum_{a,p<0}f_{-s}(a,p).
\end{align*}
Hence we obtain
\begin{align*}
h_s
&=\sum_{a>0 \atop p\geq0}f_s(a,p)+\sum_{p\geq0}f_s(0,p)-\sum_{a,p<0}f_s(a,p)\\
&=q^{-sm}\sum_{a,p<0}f_{-s}(a,p)+\sum_{p\geq0}f_s(0,p)-\sum_{a,p<0}f_s(a,p).
\end{align*}
We then have
\begin{equation}
q^{\frac12sm}h_s=q^{-\frac12sm}\sum_{a,p<0}f_{-s}(a,p)+q^{\frac12sm}\sum_{p\geq0}f_s(0,p)-q^{\frac12sm}\sum_{a,p<0}f_s(a,p).
\label{eq:1.8}
\end{equation}
A straightforward computation replacing $s$ by $-s$ yields 
\begin{equation}
q^{-\frac12sm}h_{-s}=q^{\frac12sm}\sum_{a,p<0}f_s(a,p)+q^{-\frac12sm}\sum_{p\geq0}f_{-s}(0,p)-q^{-\frac12sm}\sum_{a,p<0}f_{-s}(a,p).
\label{eq:1.9}
\end{equation}
By (\ref{eq:1.8}) and (\ref{eq:1.9}), we obtain 
\begin{align}
\label{eq:1.10}
q^{\frac12sm}h_s+q^{-\frac12sm}h_{-s}
&=2\sum_{p\geq0}q^{\frac12p(p+1)}.
\end{align}
It is known (see e.g. \S 5 in \cite{KW1}) that 
the right side of (\ref{eq:1.10}) has the following product expansion:
\[
 \sum_{p\geq0}q^{\frac12p(p+1)}=\prod_{k\geq1}\frac{1-q^{2k}}{1-q^{2k-1}}=\varphi(q)\prod_{i=1}^{\infty}(1+q^i)^2,
\]
and hence (a') follows.

Next we shall prove (b').\\
Since $f_s(-a-1,0)=-f_s(a,0)$, we have 
\[
\sum_{a\geq0}f_s(a,0)=\sum_{a<0}f_s(-a-1,0)=-\sum_{a<0}f_s(a,0).
\]
Hence we obtain
\begin{align*}
h_s&=\sum_{a\geq0 \atop p>0}f_s(a,p)
+\sum_{a\geq0}f_s(a,0)-\sum_{a,p<0}f_s(a,p)\\
&=\sum_{a\geq0 \atop p>0}f_s(a,p)
-\sum_{a<0}f_s(a,0)-\sum_{a,p<0}f_s(a,p)\\
&=(\sum_{a\geq0 \atop p>0}-\sum_{a<0 \atop p\leq0})f_s(a,p).
\end{align*}
Replacing $a$ by $-(a+1)$ and $p$ by $-p$, we have 
\begin{align}
h_s&=(\sum_{a,p<0}-\sum_{a,p\geq 0})f_s(-a-1,-p)
\nonumber \\
&=(\sum_{a,p<0}-\sum_{a,p\geq
 0})(-1)^{a+1}q^{(-p+s)(-p+s+1)-\frac{sm}2+\frac m2a(a+1)+m(a+1)p}
\nonumber \\
&=(\sum_{a,p\geq 0}-\sum_{a,p<0})(-1)^a
q^{\frac12(p-s)(p-s-1)-\frac{sm}2+\frac m2a(a+1)+m(a+1)p}.
\label{eq:1.11}
\end{align}
Replacing $s$ by $m-1-s$ in (\ref{eq:1.11})
 and comparing it with (\ref{eq:1.7}), we get (b').
\end{proof}

\vspace{2mm}
Using Lemma \ref{lem:1.1} inductively, we have the following.
\begin{proposition}
\label{prop:1.2}
Let $m\geq 2$ and $k\in \mb Z_+$. Then we have 
\begin{align*}
\mathcal{F}(e^{-\Lambda_0}ch F_{(k+1)(m-1)})&=
\mathcal{F}(e^{-\Lambda_0}ch F_{-k(m-1)})\\
&=q^{k\frac{m(m-1)}2}\big(\sum_{|j|\leq k}(-1)^{k-j}q^{(k^2-j^2)\frac{m(m-1)}2}\big)\frac{\prod^{\infty}_{i=1}(1+q^i)^2}{\varphi(q^m)^2}
\ .
\end{align*}
\end{proposition}
\begin{proof} \quad 
By Lemma \ref{lem:1.1}, we get the recurrence formula
\begin{equation}
\label{eq:1.12}
\mathcal{F}(e^{-\Lambda_0}ch F_{s+m-1})=q^{\frac{sm}2}
\left(\frac{2\prod^{\infty}_{i=1}(1+q^i)^2}{\varphi(q^m)^2}-q^{\frac{sm}2}\mathcal{F}(e^{-\Lambda_0}ch F_s)\right).
\end{equation}
We now set for $k\in \mb Z$,  $s=k(m-1)$ , $x=q^{\frac{m(m-1)}2}$  and 
\[
H_k=\mathcal{F}(e^{-\Lambda_0}ch F_{k(m-1)}){\left(\frac{\prod^{\infty}_{i=1}(1+q^i)^2}{\varphi(q^m)^2}\right)}^{-1}.\]
For the proof of this proposition, it is sufficient to show that
\begin{equation}
\label{eq:1.13}
H_{k+1}=x^k\biggl(\sum_{|j|\leq k}(-1)^{k-j}x^{k^2-j^2}\biggr),\quad k\in \mb Z_+\  .
\end{equation}
We shall show (\ref{eq:1.13}) by induction on $k$.
Using these notation, (\ref{eq:1.12}) is rewritten as 
\[
H_{k+1}=x^k(2-x^kH_k).
\]
Assume that (\ref{eq:1.13}) is true for $k-1$. Then we obtain
\begin{align*}
H_{k+1}&=x^k\big\{2-x^k\cdot
 x^{k-1}\big(\sum_{|j|\leq k-1}(-1)^{k-1-j}x^{(k-1)^2-j^2}\big)\big\}\\
&=x^k\big(2+\sum_{|j|\leq k-1}(-1)^{k-j}x^{k^2-j^2}\big)\\
&=x^k\big(\sum_{|j|\leq k}(-1)^{k-j}x^{k^2-j^2}\big),
\end{align*}
proving the proposition.
\end{proof}
We consider principally specialized characters of
$\widehat{\mf{sl}}(m|1)$-modules $L(\Lambda_{(s)})$
where $\Lambda_{(s)}$ are defined in Remark 3.2 of \cite{KW2},
given by 
\begin{equation}
\label{eq:1.14}
\Lambda_{(s)}=\begin{cases}
\Lambda_s & \hbox{if } 0\leq s\leq m\\
-(s-m)\Lambda_0+(1+s-m)\Lambda_m & \hbox{if } s\geq m\\
(1-s)\Lambda_0+s\Lambda_m+s\delta & \hbox{if }s\leq 0. 
\end{cases} \, 
\end{equation}

\begin{theorem}
\label{th:1.3}
Let $m\geq2$. Then we have the following{\rm :}
\alphaparenlist
\begin{enumerate}
\item For $\Lambda=\Lambda_0$ or $\Lambda_{m-1}$, we have 
\begin{equation}
\mathcal{F}(e^{-\Lambda}ch L(\Lambda))=\frac{\prod^{\infty}_{i=1}(1+q^i)^2}{\varphi(q^m)}.
\end{equation}
\item For $\Lambda=\{k(m-1)+1\}\Lambda_0-k(m-1)\Lambda_m$ $(k\in \mb Z)$, 
we have 
\begin{equation}
\mathcal{F}(e^{-\Lambda}ch L(\Lambda))\!=\!
\big(\sum_{|j|\leq |k|}(-1)^{k-j}q^{(k^2-j^2)\frac{m(m-1)}2}\big)\frac{\prod^{\infty}_{i=1}(1+q^i)^2}{\varphi(q^m)}.
\end{equation}
\end{enumerate}
\end{theorem}
\begin{proof} \quad 
To prove (a), we let $k=0$ in Proposition \ref{prop:1.2}.
Then, since $ch
F_s=\varphi(e^{-\delta})^{-1}ch L(\Lambda_{(s)})$ and
$\Lambda_{(0)}=\Lambda_0$, 
we get 
\begin{equation*}
\mathcal{F}(e^{-\Lambda_0}ch L(\Lambda_0))=\frac{\prod^{\infty}_{i=1}(1+q^i)^2}{\varphi(q^m)}.
\end{equation*}

Since $\Lambda_{(m-1)}=\Lambda_{m-1}=\Lambda_0-\alpha_m$, we have
\[
\mathcal{F}(e^{-\Lambda_{(m-1)}})=\mathcal{F}(e^{-\Lambda_0})\mathcal{F}(e^{\alpha_m})=\mathcal{F}(e^{-\Lambda_0}),
\]
and hence
\begin{align*}
 \mathcal{F}(e^{-\Lambda_{m-1}}ch
 L(\Lambda_{m-1}))&=\mathcal{F}(e^{-\Lambda_0}ch L(\Lambda_{(m-1)}))\\
&=\mathcal{F}(e^{-\Lambda_0}ch F_{m-1})\varphi(q^m)\\
&=\frac{\prod^{\infty}_{i=1}(1+q^i)^2}{\varphi(q^m)}
\end{align*}
which is (a).

To prove (b), we let $\Lambda=\{k(m-1)+1\}\Lambda_0-k(m-1)\Lambda_m$ for
$k\in \mb Z$.
First we consider in the case $k\in \mb Z_+$, 
since $\Lambda_m=\Lambda_0-\frac1{m-1}\alpha_1-\cdots-\frac
m{m-1}\alpha_m$ and by (\ref{eq:1.14}), 
it follows that
\[
\mathcal{F}(e^{-\Lambda_{(-k(m-1))}})=\mathcal{F}(e^{-\Lambda})q^{-km(m-1)}
=\mathcal{F}(e^{-\Lambda_0})q^{-\frac{km(m-1)}2},
\]
and hence we obtain
\begin{align*}
\mathcal{F}(e^{-\Lambda}ch L(\Lambda))
&=\mathcal{F}(e^{-\Lambda_{(-k(m-1))}}ch L(\Lambda_{(-k(m-1))}))\\
&=q^{-\frac{km(m-1)}2}
\mathcal{F}(e^{-\Lambda_0}ch L(\Lambda_{(-k(m-1))}))\\
&=\big(\sum_{|j|\leq k}(-1)^{k-j}q^{(k^2-j^2)\frac{m(m-1)}2}\big)\frac{\prod^{\infty}_{i=1}(1+q^i)^2}{\varphi(q^m)}
\ .
\end{align*}
In the case $k<0$, the same discussion as above yields
\[
\mathcal{F}(e^{-\Lambda_{((-k+1)(m-1))}})=\mathcal{F}(e^{-\Lambda})
=\mathcal{F}(e^{-\Lambda_0})q^{\frac{km(m-1)}2},
\]
and hence we obtain 
\begin{align*}
\mathcal{F}(e^{-\Lambda}ch L(\Lambda))
&=\mathcal{F}(e^{-\Lambda_{((-k+1)(m-1))}}ch L(\Lambda_{((-k+1)(m-1))}))\\
&=q^{\frac{km(m-1)}2}
\mathcal{F}(e^{-\Lambda_{0}}ch L(\Lambda_{((-k+1)(m-1))}))\\
&=\big(\sum_{|j|\leq
 -k}(-1)^{k-j}q^{(k^2-j^2)\frac{m(m-1)}2}\big)\frac{\prod^{\infty}_{i=1}(1+q^i)^2}{\varphi(q^m)}
\end{align*}
which is (b).
\end{proof}

\vspace{2mm}
\begin{remark}\normalfont
\quad This is an additional remark to Theorem \ref{th:1.3}.\\
We consider an asymptotic behavior 
of principally specialized characters 
$\mathcal{F}(e^{-\Lambda_0}ch L(\Lambda_0))$.

We shall write $f(\tau)\underset{\tau\downarrow0}{\sim} g(\tau)$ if
${\rm lim}_{\tau\downarrow0}f(\tau)/g(\tau)=1$, 
where $\tau\downarrow0$ means that 
$\tau=iT\ (T>0)$ and $T\rightarrow 0.$

It is known (see e.g. (13.13.5) in \cite{K3}) that the asymptotic
behavior of the Dedekind $\eta$-function
\[
\eta(\tau)=q^{\frac1{24}}\varphi(q)=q^{\frac1{24}}\prod^{\infty}_{n=1}(1-q^n)
  \quad (q=e^{2\pi i\tau})
\]
is given by
\[
\eta(\tau)\underset{\tau\downarrow0}{\sim}
(-i\tau)^{-\frac12}e^{-\frac{\pi i}{12\tau}},
\]
and so 
\[
\varphi(q)=q^{-\frac1{24}}\eta(\tau)
\underset{\tau\downarrow0}{\sim}(-i\tau)^{-\frac12}e^{-\frac{\pi i}{12\tau}}.
\]
>From this assertion, we have 
\[
\prod^{\infty}_{n=1}(1+q^n)=
\frac{\varphi(q^2)}{\varphi(q)}
\underset{\tau\downarrow0}{\sim}\frac1{\sqrt2}e^{\frac{\pi i}{24\tau}}
\]
and hence
\begin{eqnarray*}
\mathcal{F}(e^{-\Lambda_0}ch L(\Lambda_0))
&\underset{\tau\downarrow0}{\sim}&
\frac{\frac12e^{\frac{\pi i}{12\tau}}}
{(-im\tau)^{-\frac12}e^{-\frac{\pi i}{12m\tau}}}\\
&=&\frac{\sqrt m}2(-i\tau)^{\frac12}e^{\frac{2\pi i}{\tau}\frac{m+1}{24m}}.
\end{eqnarray*}

By applying the Tauberian theorem(see \cite{I} and Proposition 4.22 in
\cite{KP}), we have 
\[
a_n\ \underset{n\rightarrow\infty}{\sim}\ 
\frac1{8\sqrt3\ n}(m+1)^{\frac12}e^{\pi\sqrt{\frac23\frac{m+1}mn}}
\]
where 
$\mathcal{F}(e^{-\Lambda_0}ch L(\Lambda_0))
=\sum^{\infty}_{n=0}a_nq^n$.

Next, we consider for principally specialized characters 
of some series of level~1 integrable modules.
By Theorem \ref{th:1.3}, we have for  $\Lambda=\Lambda_{m-1}$ or 
$\{k(m-1)+1\}\Lambda_0-k(m-1)\Lambda_m$ $(k\in \mb Z)$,
\begin{align}
\mathcal{F}(e^{-\Lambda}ch L(\Lambda))
&\underset{\tau\downarrow0}{\sim}
\mathcal{F}(e^{-\Lambda_0}ch L(\Lambda_0))
\nonumber \\
&\underset{\tau\downarrow0}{\sim}
\frac{\sqrt m}2(-i\tau)^{\frac12}e^{\frac{2\pi i}{\tau}\frac{m+1}{24m}}.
\label{eq:1.17}
\end{align}

The formula (4.12) in \cite{KW2} is the asymptotics of the basic specialization 
(the specialization of type $(1,0,\ldots ,0)$)
of $\widehat{\mf{sl}}(m|1)$-module $L(\Lambda_{(s)})$.
The formula which is obtained by replaced $\tau$ by $m\tau$ in (4.12) of \cite{KW2}
coincides with (\ref{eq:1.17}).
\end{remark}

%%%%%%%%%%%%%%%%%%%%%%%%%%%%%%%%%%%%%%%%%%%%%%%%%%%%%%%%%%%%%%%%%%%%%%%%%
%%%%%%%%%%%%%%%%%%%%%%%%%%%%%%%%%%%%%%%%%%%%%%%%%%%%%%%%%%%%%%%%%%%%%%%%%
%%%%%%%%%%%%%%%%%%%%%%%%%%%%%%%%%%%%%%%%%%%%%%%%%%%%%%%%%%%%%%%%%%%%%%%%%
\section{Specialized character formula of ``quasiparticle'' type}
\label{sec:2}

\begin{proposition}
\label{prop:2.1}
For $m\geq 2$ and $s\in \mb Z$, we have
\[
\mathcal{F}(e^{-\Lambda_0} ch F_s)=q^{-\frac{sm}2}\sum_{a,b,c,d\in{\mb Z}_+\atop a-b+c-d=s}
\frac{q^\frac{a(a+1)}2q^\frac{b(b-1)}2q^{cm}}{(q)_a(q)_b(q^m)_c(q^m)_d}\ .
\]
Here and further $(q)_n=\prod^n_{j=1}(1-q^j)$.
\end{proposition}
\begin{proof} \quad 
This formula is shown just by the same argument as that in the proof of the formula
(3.14) in \cite{KW2}, by making use of a basis of $F_s$:
\[
\begin{array}{l}
(\psi^{(1)}_{-(j_{1,1}-\frac12)}\cdots\psi^{(1)}_{-(j_{1,a_1}-\frac12)})
\cdots
(\psi^{(m)}_{-(j_{m,1}-\frac12)}\cdots\psi^{(m)}_{-(j_{m,a_m}-\frac12)})
\\
\times 
(\psi^{(1)*}_{-(j'_{1,1}-\frac12)}\cdots\psi^{(1)*}_{-(j'_{1,b_1}-\frac12)})
\cdots
(\psi^{(m)*}_{-(j'_{m,1}-\frac12)}\cdots\psi^{(m)*}_{-(j'_{m,b_m}-\frac12)})
\\
\times 
(\varphi^{(1)}_{-(k_1-\frac12)}\cdots\varphi^{(1)}_{-(k_c-\frac12)})
(\varphi^{(1)*}_{-(k'_1-\frac12)}\cdots\varphi^{(1)*}_{-(k'_d-\frac12)})
|0\rangle
\\
\end{array}
\]
satisfying
\[
\begin{array}{l}
0<j_{i,1}<\cdots<j_{i,a_i},0<j'_{i,1}<\cdots<j'_{i,b_i}\ \ \ (i=1,\ldots,m),
\\
0<k_1\leq\cdots\leq k_c,0<k'_1\leq\cdots\leq k'_d
\end{array}
\]
and
\[
a-b+c-d=s 
\]
where $a=\sum_ia_i,b=\sum_ib_i$.

Since
\begin{align}
{\rm weight} (\psi^{(i)}_k) &= \epsilon_i + k \delta,\qquad \ 
{\rm weight} (\psi^{(i)*}_k) = -\epsilon_i + k \delta, 
\nonumber \\
{\rm weight} (\varphi^{(1)}_k) &= \epsilon_{m+1} + k \delta ,\quad 
{\rm weight} (\varphi^{(1)*}_k) = -\epsilon_{m+1} +k \delta ,
\nonumber 
\end{align}
and (\ref{eq:1.2}), we have
\[
\begin{array}{ll}
\mathcal{F}\big(\ e^{{\rm weight} (\psi^{(i)}_{-(j-\frac12)})}\ \big)
=q^{i+m(j-\frac32)},
&\mathcal{F}\big(\ e^{{\rm weight} (\psi^{(i)*}_{-(j-\frac12)})}\ \big)
=q^{-i+m(j+\frac12)},\\
\mathcal{F}\big(\ e^{{\rm weight} (\varphi^{(1)}_{-(j-\frac12)})}\ \big)
=q^{m(j-\frac12)},
&\mathcal{F}\big(\ e^{{\rm weight} (\varphi^{(1)*}_{-(j-\frac12)})}\ \big)
=q^{m(j-\frac12)},
\end{array}
\]
for $i=1,\ldots ,m$ and $j\in \mb N$.

\vspace{3mm}
Hence we have
\vspace{-1mm}
\begin{align}
\mathcal{F}(e^{-\Lambda_0} ch F_s)=&\!\!\!\!
\sum_{a,b,c,d\in{\mb Z}_+\atop a-b+c-d=s}\!\!
\bigg(\sum_{0<i_1<\cdots <i_a}\!\!\!q^{(i_1-\frac m2)+\cdots +(i_a-\frac m2)}\!\bigg)
\bigg(\sum_{0<j_1<\cdots <j_b}\!\!\!q^{(j_1+\frac m2-1)
+\cdots +(j_b+\frac m2-1)}\!\bigg)
\nonumber \\
&\quad \times \bigg(\sum_{0<k_1\leq \cdots \leq k_c}\!\!\!
q^{(mk_1-\frac m2)+\cdots +(mk_c-\frac m2)}\!\bigg)
\bigg(\sum_{0<l_1\leq \cdots \leq l_d}\!\!\!q^{(ml_1-\frac m2)
+\cdots +(ml_d-\frac m2)}\!\bigg)
\nonumber \\
=&q^{-\frac{sm}2}\sum_{a,b,c,d\in{\mb Z}_+\atop a-b+c-d=s}
\bigg(\sum_{0<i_1<\cdots <i_a}q^{i_1+\cdots +i_a}\bigg)
\bigg(q^{-b}\sum_{0<j_1<\cdots <j_b}q^{j_1+\cdots +j_b}\bigg)
\nonumber \\
&\quad \times \bigg(\sum_{0<k_1\leq \cdots \leq k_c}q^{m(k_1+\cdots +k_c)}\bigg)
\bigg(q^{-dm}\sum_{0<l_1\leq \cdots \leq l_d}q^{m(l_1+\cdots +l_d)}\bigg)
\nonumber \\
=&q^{-\frac{sm}2}\sum_{a,b,c,d\in{\mb Z}_+\atop a-b+c-d=s}\frac{q^{\frac12a(a+1)}}{(q)_a}\frac{q^{\frac12b(b-1)}}{(q)_b}
\frac{q^{cm}}{(q^m)_c}\frac1{(q^m)_d} \ ,
\nonumber 
\end{align}
proving the proposition.
\end{proof}
\vspace{4mm}
By Theorem \ref{th:1.3} and Proposition \ref{prop:2.1}, we also have shown
\begin{corollary}
\label{cor:2.2}
For $m\geq2$, the following formula holds
\[
\frac{\prod^{\infty}_{i=1}(1+q^i)^2}
{\varphi(q^m)^2}
=\sum_{a,b,c,d\in{\mb Z}_+\atop a-b+c-d=0}
\frac{q^\frac{a(a+1)}2q^\frac{b(b-1)}2q^{cm}}{(q)_a(q)_b(q^m)_c(q^m)_d}\ .
\]
\end{corollary}

%%%%%%%%%%%%%%%%%%%%%%%%%%%%%%%%%%%%%%%%%%%%%%%%%

\end{document}